\documentclass[aps, prd, letterpaper, 12pt, nofootinbib, superscriptaddress, longbibliography, notitlepage]{revtex4-1}
\usepackage[utf8]{inputenc}
\usepackage{amsmath,amssymb,amsfonts}
\usepackage{mathrsfs}
\usepackage{color}
\usepackage{graphicx} 
\usepackage[section]{placeins}
\usepackage[colorlinks=true,linkcolor=blue,citecolor=blue]{hyperref}

\allowdisplaybreaks

\usepackage{tikz}
\usepackage{tkz-euclide}
\usetikzlibrary{decorations.pathmorphing}	
\tikzset{
    v/.style={decorate, decoration={snake, segment length=3mm, amplitude=0.75mm}, draw},
    f/.style={draw,decoration={markings,mark=at position #1 with {\arrow[very thick]{latex}}},postaction={decorate},node contents=#1},
    f/.default=.6,
    fb/.style={draw,decoration={markings,mark=at position #1 with {\arrowreversed[very thick]{latex}}},postaction={decorate},node contents=#1},
    fb/.default=.4,
    fnar/.style={draw},
    g/.style={decorate, draw,  decoration={coil,amplitude=3pt, segment length=3.5pt}},
    s/.style={dashed,draw, postaction={decorate},
        decoration={markings,mark=at position .55 with {\arrow[very thick]{latex}}}},
    sb/.style={dashed,draw, postaction={decorate},
        decoration={markings,mark=at position .55 with {\arrowreversed[draw=black,very thick]{latex}}}},
    snar/.style={dashed,draw,line width =1.25pt},
}
\usetikzlibrary{shapes}	
\tikzset{every picture/.style={line width=1}}

\definecolor{c1}{rgb}{1., 0.49804, 0.05492}
\definecolor{c2}{rgb}{0.83922, 0.15294, 0.15686}
\definecolor{c3}{rgb}{0.12157, 0.46667, 0.70588}
\definecolor{c4}{rgb}{0.73725, 0.74118, 0.13333}
\definecolor{c5}{rgb}{0.17647, 0.62745, 0.17647}
\definecolor{c6}{rgb}{0.89020, 0.47059, 0.76078}
\definecolor{c7}{rgb}{0.58039, 0.40392, 0.74118}

\begin{document}

\title{Astrophysical constraints from synchrotron emission\\ on very massive decaying dark matter}

\author{Pankaj~Munbodh}
\email{pmunbodh@ucsc.edu}
\affiliation{Department of Physics, University of California Santa Cruz, and
Santa Cruz Institute for Particle Physics, 1156 High St., Santa Cruz, CA 95064, USA}

\author{Stefano~Profumo}
\email{profumo@ucsc.edu}
\affiliation{Department of Physics, University of California Santa Cruz, and
Santa Cruz Institute for Particle Physics, 1156 High St., Santa Cruz, CA 95064, USA}

\begin{abstract}
\noindent If the cosmological dark matter (DM) couples to Standard Model (SM) fields, it can decay promptly to SM states in a highly energetic hard process, which subsequently showers and hadronizes to give stable particles including $e^\pm$, $\gamma$, $p^{\pm}$ and $\nu\bar{\nu}$ at lower energy. If the DM particle is very heavy, the high-energy $e^\pm$, due to the Klein-Nishina cross section suppression, preferentially lose energy via synchrotron emission which, in turn, can be of unusually high energies.  Here, we present previously unexplored bounds on heavy decaying DM up to the Planck scale, by studying the synchrotron emission from the $e^\pm$ produced in the ambient Galactic magnetic field. In particular, we explore the sensitivity of the resulting constraints on the DM decay width to (i) different SM decay channels, to (ii) the Galactic magnetic field configurations, and (iii) to various different DM density profiles proposed in the literature. We find that constraints from the synchrotron component complement and improve on constraints from very high-energy cosmic-ray and gamma-ray observatories targeting the prompt emission when the DM is sufficiently massive, most significantly for masses in excess of $10^{12}\text{ GeV}$.
\end{abstract}

\maketitle

\newpage

\section{Introduction} \label{sec:intro}
Observations of the dynamics of galaxy clusters, galactic rotation curves, the Cosmic Microwave Background (CMB) and large scale structure provide strong evidence for the existence of Dark Matter (DM). In particular, DM must be stable on cosmological time scales \cite{Poulin:2016nat, Audren:2014bca}. 
If the DM is a particle that interacts with the Standard Model (SM), its possible decay to SM states allows us to place even stronger bounds on its lifetime. If the DM mass $m_\chi$ is large, the decay process $\chi \to \text{SM+SM}$ produces the SM states with an energy $\mu \sim m_\chi/2$; such states subsequently hadronize and shower to produce photons, electrons/positrons, protons/anti-protons and neutrinos below the electroweak (EW) scale \cite{Bauer:2020jay}.

The effort in searching for photons from DM annihilation and/or decay, part of the broader multi-messenger ``indirect DM detection'' campaign, is at a very advanced stage (for a pedagogical review, see \cite{Slatyer:2021qgc}). In particular, the leading facilities probing the DM particle lifetime indirectly through its final decay products, such as photons, neutrinos and cosmic rays, include the Fermi Large Area Telescope (Fermi-LAT) \cite{Huang:2011xr, Song:2023xdk, Fermi-LAT:2012edv}, the Pierre Auger Observatory (PAO) \cite{PierreAuger:2015eyc, Aloisio:2022eqx}, the High-Altitude Water Cherenkov Observatory (HAWC) \cite{HAWC:2017udy, Albert:2023ucd}, the Large
High Altitude Air Shower Observatory (LHAASO) \cite{LHAASO:2022yxw, LHAASO:2019qtb} and IceCube \cite{IceCube:2018tkk, IceCube:2023gku, Esmaili:2013gha, Chianese:2019kyl}. In addition, upcoming experiments including the Cherenkov Telescope Array (CTA) \cite{CTAConsortium:2017dvg,Pierre:2014tra, CTA:2020qlo} and IceCube-Gen2 \cite{IceCube-Gen2:2020qha,Fiorillo:2023clw, Murase:2015gea} are expected to ramp up the reach of future indirect DM detection campaigns. 

The products of the prompt decay of DM into SM particles are not exhaustive of the entirety of the emission eventually arising from the decay event. Stable, charged particles, electrons and positrons ($e^\pm$) and protons and antiprotons, lose energy to a variety of electromagnetic processes that, in turn, produces lower-energy radiation. Due to the difference in particle mass, the emission from $e^\pm$ is both brighter and higher-energy than that from hadrons, albeit it does not include inelastic processes. As a function of the $e^\pm$ energy, from low to high energy, the principal energy loss mechanisms are, typically, Coulomb losses, bremsstrahlung, Inverse Compton (IC), and synchrotron \cite{Longair:1992ze}. The first detailed calculation of the broad band emission from a specific class of weakly-interacting massive particles (WIMPs), supersymmetric neutralinos, was carried out in Ref.~\cite{Colafrancesco:2005ji, Colafrancesco:2006he} (see also \cite{Profumo:2010ya} for a review). Since then, several other studies have investigated the prompt emission, Inverse Compton emission and synchrotron emission for WIMP-like, electroweak-scale DM masses (an incomplete list includes Ref.~\cite{Hutsi:2010ai,Cohen:2016uyg, Blanco:2018esa}), while other studies have explored the prompt and IC emission for superheavy DM \cite{Esmaili:2015xpa, Kalashev:2020hqc,Chianese:2021jke, Esmaili:2021yaw, Chianese:2021htv, Skrzypek:2021tuv, Maity:2021umk}\footnote{Note that Ref. \cite{Skrzypek:2021tuv} includes synchrotron energy losses, but not emission.}. Ref. \cite{Deliyergiyev:2022bvp} derives constraints on superheavy DM combining prompt emission with gravitational wave observations. 

In this paper, to our knowledge for the first time, we focus on bounds obtained from {\em synchrotron emission} emanating from superheavy DM decays. The best observational target to search for this signal is by far the Galactic center. Here, we intend to assess whether the synchrotron emission, at much lower energy than the prompt and IC emission, provides comparable or more competitive bounds on the DM lifetime. The peak frequency of the synchrotron emission depends linearly on the ambient magnetic field, and quadratically on the electron energy. In particular, we will show that the peak energy of the synchrotron emission, in the Galactic center region, scales with the DM mass $m_\chi$, approximately as
\begin{equation}
    \frac{E_{\rm peak}}{\rm GeV}\simeq \left(\frac{m_\chi}{10^{10}\ {\rm GeV}}\right)^2.
\end{equation}
As a result, the synchrotron emission falls squarely in the high-energy gamma-ray range where Fermi-LAT is sensitive (roughly between 0.1 GeV and 1 TeV) for $10^9\lesssim m_\chi/{\rm GeV}\lesssim 10^{12}$; in the very high-energy Cherenkov telescope range (roughly 0.1 TeV to 1,000 TeV) for  $10^{11}\lesssim m_\chi/{\rm GeV}\lesssim 10^{14}$; and, at even larger masses, at ultra high-energy cosmic-ray/gamma-ray facilities.

The principal goal of the present study is to understand in detail how the predicted synchrotron emission depends on the assumed Galactic magnetic model, on the DM density distribution in the Galaxy, on the $e^\pm$ injection spectrum, and the signal morphology. To this end, we first review, in Section \ref{sec:theory}, the calculation of the prompt gamma-ray emission , the $e^\pm$ injection spectrum, the subsequent energy losses, and the resulting synchrotron emission. In Section \ref{sec:analysis}, we elaborate on the dependence of the signal on the Galactic magnetic field model and on the DM density profiles that we implement. In Section \ref{sec:exp_lim}, we briefly describe the experimental limits set by astrophysical multimessenger experiments (Fermi-LAT, HESS \cite{HESS:2022ygk}, Pierre-Auger Observatory, CASA-MIA~\cite{CASA-MIA:1997tns}, KASCADE ~\cite{KASCADEGrande:2017vwf}, KASCADE-GRANDE~\cite{KASCADEGrande:2017vwf}, Telescope Array Surface Detector (TASD)~\cite{TelescopeArray:2018rbt} and EAS-MSU~\cite{Fomin:2017ypo}) on the observed photon flux from the Galactic center. In Section \ref{sec:results}, we compute the synchrotron fluxes for various initial SM states and we present previously unexplored constraints on the lifetime of the DM depending on its mass. Finally, we summarize our findings and present our outlook in Section \ref{sec:conclusions}.

\section{Multi-wavelength Emission From Dark Matter Decay} \label{sec:theory}
Here we review schematically the production of photons from heavy DM decay, both via prompt production (sec.~\ref{sec:prompt}) and via synchrotron emission off of the prompt $e^\pm$ (sec.~\ref{sec:sync}).
\subsection{Prompt Emission}\label{sec:prompt}
The differential flux of photons from DM decay from a given line of sight is given by \cite{Esmaili:2015xpa}
\begin{equation}
    \frac{d\Phi}{dE_\gamma} = \frac{1}{4\pi m_\chi \tau_\chi} \frac{dN}{dE_\gamma} \int_0^\infty ds\, \rho (s,b,l) e^{-\tau_{\gamma\gamma}(E_\gamma,s,b,l)}
\end{equation}
where $E_\gamma$ is the photon energy, $m_\chi$ is the DM mass, $\tau_\chi$ is the DM lifetime, $\rho$ is the Galactic DM density profile, $s$ is the line-of-sight distance from the observer, and $b,l$ are the Galactic latitude and longitude angular coordinates respectively. $dN/dE_\gamma$ is the energy spectrum of photons produced per decay and $\tau_{\gamma\gamma}$ is the optical depth due to pair production against Cosmic Microwave Background (CMB), starlight (SL) and infrared (IR) photons.

The energy spectrum $dN/dE_\gamma$ is obtained from {\tt HDMSpectra} \cite{Bauer:2020jay} which incorporates EW corrections that become important at higher energies\footnote{Some of these EW corrections such as triple gauge couplings in the EW sector are neglected by Pythia.}.

The optical depth $\tau_{\gamma\gamma}$, which characterizes the impact of the absorption of gamma rays in the interstellar medium, has a noticeable effect on the spectrum mostly in the range of energies $E_\gamma \simeq 10^4 - 10^8$ GeV. Our analysis neglects the optical depth entirely since we expect that the constraints we derive from high energy gamma-ray probes of the integrated flux (see the discussion in Section \ref{sec:exp_lim}) suffer from at most $O(1)$ corrections due to the effects of attenuation from the Galactic center to the Earth. Please see Refs. \cite{Esmaili:2021yaw, DeAngelis:2013jna} for more details about the flux attenuation. Of course this assumption is invalid for extragalactic DM decays. Attenuation and electromagnetic cascades cause washout of the primary emission spectra such that Ref.~\cite{Murase:2012xs} derives largely mass-independent constraints for the extragalactic component to the diffuse gamma ray background\footnote{In the mass range of $10^3 - 10^{10}$ GeV, Ref.~\cite{Murase:2012xs} includes both galactic and extragalactic contributions in their derived bounds. In computing the synchrotron emission, they assume a value of 1 $\mu$G for the galactic magnetic field.}.

\subsection{Synchrotron emission}\label{sec:sync}
\begin{figure}[h!]
\centering
\includegraphics[width=0.7\linewidth]{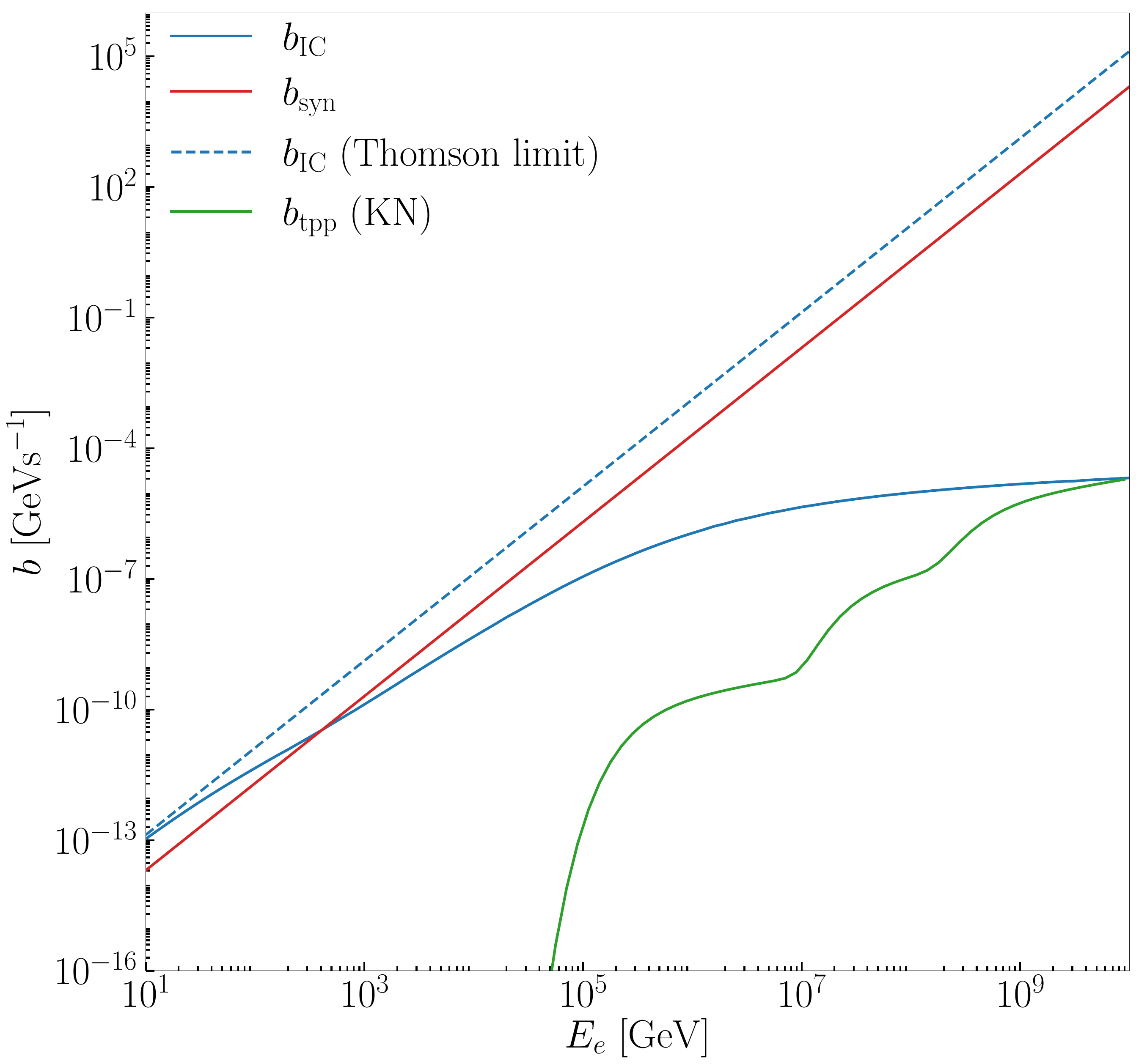}
\caption{Energy losses from synchrotron emission ($b_{\rm syn}$) from the Galactic Center, Inverse Compton emission ($b_{\rm IC}$), Inverse Compton emission in the Thomson limit ($b_{\rm IC}$ Thomson) and triplet pair production processses in the Klein-Nishina regime ($b_{tpp}$ KN). At ultrarelativistic electron energies $E_e$ close to half the DM mass $m_\chi/2$, synchrotron emission vastly dominates over IC and TPP emission.} 
\label{fig:energy}
\end{figure}

Relativistic electrons in the Galactic magnetic field produce synchrotron radiation \cite{RevModPhys.42.237}. At very high energy, typical of the $e^\pm$ produced by the decay of very massive DM, the inverse Compton Klein-Nishina cross section is highly suppressed compared to the corresponding synchrotron emission process. The differential synchrotron component of the gamma ray flux (after setting the optical depth to zero) is given by
\begin{equation}
\label{eq:synchrotron_flux}
    \frac{d\Phi}{dE_\gamma d\Omega}(E_\gamma,b,l) = \frac{1}{4\pi E_\gamma} \int_{m_e}^{m_\chi/2}dE_e \, \int_0^\infty ds\, \frac{dn_e}{dE_e}(E_e,s,b,l)P_{\rm syn}(E_\gamma, E_e,s,b,l),
\end{equation}
where $m_e$ is the mass of the electron and $d\Omega = \cos b \, db \, dl$ is the solid angle. The formula for the synchrotron power $P_{\rm syn}$ is provided in Appendix \ref{app:Psyn}. The steady-state equilibrium electron number density after energy losses and diffusion, is given by
\begin{equation}
    \frac{dn_e}{dE_e} (E_e,s,b,l) = \frac{1}{m_\chi \tau_\chi} \frac{\rho(s,b,l)}{b_{\rm tot}(E_e,s,b,l)} \int^{m_\chi/2}_{E_e} \frac{dN}{dE_e^\prime}I(E_e,E_e^\prime,s,b,l) \, dE_e^\prime,
\end{equation}
where $b_{\rm tot} = -dE_e/dt$ is the total energy loss coefficient comprising energy losses due to Inverse Compton (IC) scattering of electrons on CMB, SL and IR photons ($b_{\rm IC}$), triplet pair production (TPP) energy losses ($b_{\rm TPP}$) \cite{10.1093/mnras/253.2.235} and synchrotron energy losses ($b_{\rm syn}$). As mentioned above, for superheavy DM (which subsequently produce ultrarelativistic electrons below the EW scale), we verified that TPP and IC energy losses are negligible when compared to synchrotron i.e. $b_{\rm TPP}, b_{\rm IC} \ll b_{\rm syn}$ so that taking
\begin{equation}
    b_{\rm tot} \simeq  b_{\rm syn}
\end{equation}
is entirely sufficient for our purposes. Typically, for the photon energy ranges $E_\gamma$ that we consider, the integrand in Eq.~\eqref{eq:synchrotron_flux} reaches its maximum value when $E_e$ is within an order of magnitude of half the DM mass $m_\chi/2$ indicating that most of the synchrotron emission comes from highly energetic electrons.
The energy losses as a function of electron energy $E_e$ are illustrated in Figure \ref{fig:energy}. For $E_e \gtrsim 10^3$ GeV, synchrotron energy loss (solid red line) starts to dominate over IC (solid blue line) and Triplet Pair Production (TPP, solid green line) energy losses. For comparison, we have also shown the IC energy loss in the Thomson limit (an approximation which breaks down as $E_e$ grows) as a dotted blue line. Details about the computation of the energy losses are found in Appendix \ref{app:energy_losses}.

Similarly to the prompt photon emission, the injected energy spectrum of electrons $dN/dE_e$ is obtained from {\tt HDMSpectra}, whereas the diffusion halo function $I$ can be solved  from the diffusion-loss equation as discussed in Ref. \cite{Cirelli:2010xx} (see also \cite{Colafrancesco:2005ji}). At high injection energies, $I\simeq 1$ \cite{Cirelli:2010xx, Esmaili:2015xpa}.
Defining $x = 2 E_e/m_\chi$, we  thus have:
\begin{equation}
\label{eq:elec_density}
    \frac{dn_e}{dE_e} \simeq \frac{1}{m_\chi \tau_\chi} \frac{\rho(s,b,l)}{b_{\rm tot}(E_e,s,b,l)} \int^1_{2\frac{E_e}{m_\chi}} \frac{dN}{dx}\, dx.
\end{equation}
Replacing the electron number density from equation \eqref{eq:elec_density} and $P_{\rm syn}$ from equation \eqref{eq:Psyn} into equation \eqref{eq:synchrotron_flux}, we are able to find the differential flux in a given angular region along a certain direction ($b,l$) in the sky.

For reference, we also list the angular averaged flux over the unit sphere $S^2$ given by
\begin{equation}
\label{eq:synchrotron_flux_avg}
    \frac{d\Phi}{dE_\gamma}(E_\gamma)\Bigg |_{S^2 \text{ avg}} = \frac{1}{4\pi} \int_{S^2} d\Omega \frac{d\Phi}{dE_\gamma d\Omega}
\end{equation}
and the angular averaged $E_\gamma$ integrated flux,
\begin{equation}
\label{eq:synchrotron_flux_int}
    \Phi(E_\gamma) = \frac{1}{4\pi} \int_{S^2} d\Omega \int_{E_\gamma^\prime}^{\frac{m_\chi}{2}} dE_\gamma^\prime \frac{d\Phi}{d\Omega dE_\gamma^\prime}
\end{equation}
Both of these quantities will be useful during the discussion in Section \ref{sec:exp_lim}.

\section{Astrophysical inputs: Magnetic field and DM density}
\label{sec:analysis}
In this Section, we summarize the Galactic magnetic field models (in \ref{subsec:mag_field}) and DM density profiles (in \ref{subsec:DM_profile}) implemented in our study. 

\subsection{Magnetic field models}
\label{subsec:mag_field}
The morphology of the Galactic magnetic field (GMF) is highly uncertain due to the lack of direct measurements \cite{Jansson:2012pc, Unger:2017wyw}. A number of GMF models have been proposed in the literature with different functional forms \cite{Sun:2007mx, Sun:2010sm, Jaffe:2013yi}. For recent reviews on the GMF and modeling efforts, see also Refs. \cite{Jaffe:2019iuk,refId0}.

Here, we employ the Jansson and Farrar (JF12) magnetic field model as our benchmark \cite{Jansson:2012pc, Jansson:2012rt}. This model uses the
WMAP7 Galactic Synchrotron Emission map and more than forty thousand extragalactic rotation measures to fit the model parameters to observations. JF12 consists of a disk field and an extended halo field, both containing large scale regular field, small scale random field and striated random field components.
Detailed information about the implementation of the JF12 model and its parameters can be found in Refs. \cite{Jansson:2012pc, Jansson:2012rt}. During the preparation of this work, we became aware of Ref. \cite{Unger:2023lob} which analyses variants of the coherent disk, poloidal and toroidal halo fields, free from discontinuities, together with variants for the thermal and cosmic-ray electron densities, to finally converge to an optimized set of eight fitted GMF models. The parametric models employed in the fitting procedure are more complex but many similarities regarding the overall structure of the GMF models are shared with JF12\footnote{See the discussion and Figures appearing in Section 7 of Ref~\cite{Unger:2023lob}.}. Most importantly, the magnetic field strength of the fitted ensemble are of the same order as JF12. Therefore, we expect that our results are largely insensitive and qualitatively unaffected by the updated model in Ref.~\cite{Unger:2023lob}.

Instead of realizing a stochastic implementation of the random and random striated components, we adopt the estimate discussed in \cite{Jansson:2012pc, Jansson:2012rt},  i.e. we approximate the rms value of the striated component by $B^2_{\rm stri} = \beta B^2_{\rm reg}$, whereas the relativistic electron density is rescaled by a factor $n_e \to \alpha n_e$, where $\alpha$ and $\beta$ are parameters of the fitted JF12 model.
Hence in our formula for $b_{\rm syn}$, we calculate $B$ using the prescription
\begin{equation}
\label{eq:B_prescription1}
    B = \sqrt{(\beta + 1)B_{\rm reg}^2 + B_{\rm ran}^2}
\end{equation}
where $B_{\rm ran}$ is the rms value of the random component.
Since the synchrotron flux is proportional to $n_e$ and the perpendicular component (squared) $B_{\perp}^2 = B^2\sin^2 \theta$, we make the replacement
\begin{equation}
\label{eq:B_prescription2}
    B_{\perp}^2 \to \alpha(1+\beta)B^2_{\rm reg}\sin^2\theta + \frac{2}{3}\alpha B^2_{\rm ran} 
\end{equation}
in $P_{\rm syn}$ where the factor of $(2/3)$ comes from averaging isotropically. Equations \eqref{eq:B_prescription1} and \eqref{eq:B_prescription2} suffice to provide an accurate estimate of the synchrotron emission using the JF12 model, as we verified by comparing the assumed synchrotron energy losses with the synchrotron emission integrated over energy.

Ref. \cite{Adam2016PlanckIR} gives updated parameters for the JF12 model by including information about the total and polarized dust emission, as well as synchrotron at low frequency (30 GHz) from Planck data. In particular, the Jansson 12b (JF12b) model is adjusted to match the synchrotron emission\footnote{JF12b underestimates the dust polarization at high latitudes.} while the Jansson 12c (JF12c) model is adjusted to match the dust emission while trying to retain the features of JF12b. We further emphasize that neither JF12b nor JF12c give best-fit results to the data but only indicate how the original JF12 model parameters could be improved. 

Finally, due to its simplicity, we also implement the MF1 model as defined in Ref.~\cite{Cirelli:2010xx} (see also Refs.~\cite{Perelstein:2010gt,Strong:1999sv}). 
MF1 only consists of a regular component for the disk field
\begin{equation}
    B(s,b,l) = B_0 \exp\left [{-\frac{|r_{\rm cyl} - R_\odot|}{r_B} - \frac{|z|}{z_B}}\right ]
\end{equation}
where $r_{\rm cyl} = \sqrt{s^2 \cos^2 (b) + R_{\odot}^2 - 2 s R_{\odot}\cos(b)\cos(l)}$ is the galactocentric radius, $z = s \sin (b)$,  $R_\odot = 8.3 \text{ kpc}$, $r_B = 10 \text{ kpc}$, $z_B = 2 \text{ kpc}$ and $B_0 = 4.78~ \mu G$. We caution that MF1 is an  over-simplified model of the GMF as it does not contain a halo field\footnote{ The local halo field strength is known to exceed 1.6 $\mu$G \cite{10.1093/mnras/stz1060}.} and additionally fails to capture any turbulent components of the GMF. Since we aim to compare our results from JF12 with arguably the simplest GMF model, we refrain from augmenting MF1 with halo field or turbulent field components.

\subsection{DM density profiles}
\label{subsec:DM_profile}
The determination of the distribution of DM at the Galactic center suffers from a large degree of uncertainty due to the axisymmetric Galactic bar and non-circular streaming motions \cite{Sofue:2013kja}. 
Therefore, we find it prudent to explore the sensitivity of our results to different DM density profiles which are summarized below.

We choose as our benchmark the standard Navarro–Frenk–White (NFW) profile for the DM density \cite{Navarro:1995iw, Navarro:1996gj}.
\begin{equation}
    \rho_{\rm NFW}(r) = \frac{\rho_h}{\frac{r}{r_s}\left (1+\frac{r}{r_s}\right )^2}
\end{equation}
where $r$ is the spherical coordinate
\begin{equation}
    r = \sqrt{s^2 + R{_\odot}^2 - 2sR_{\odot}\cos(b)\cos(l)}
\end{equation}
and $\rho_h = 0.18 \text{ GeVcm}^{-3}$, $r_s = 24 \text{ kpc}$ is the scale radius.
 
In addition to the NFW profile, we also consider its generalized form (gNFW), given by
\begin{equation}
    \rho_{\rm gNFW}(r) = \frac{M_0}{4\pi r_s^3}\frac{1}{\left (\frac{r}{r_s}\right )^\beta \left ( 1+ \frac{r}{r_s}\right )^{3-\beta}}
\end{equation}
Ref. \cite{Lim:2023lss} gives the best-fit values $M_0 = 1.3778\times 10^{11} M_\odot$, $r_s = 3.6$ kpc and $\beta = 1.1$, whereas Ref. \cite{Ou:2023adg} gives $ M_0 =   3.21 \times 10^{11} M_\odot$, $r_s = 5.26 \text{ kpc}$ and $\beta = 0.0258$. We will refer to the first set of parameter values as gNFWLim and the second set as gNFWOu.

Finally, we also consider the Einasto (EIN) profile
\begin{equation}
    \rho_{\rm Ein}(r) = \frac{M_0}{4\pi r_s^3}e^{-(r/r_s)^\alpha};
\end{equation}
For the EIN input parameters, Ref. \cite{Ou:2023adg} finds $M_0 = 0.62 \times 10^{11} M_\odot$, $r_s = 3.86 \text{ kpc}$ and $\alpha = 0.91$. We note that Ref.~\cite{Ou:2023adg} argues that the Einasto profile is statistically preferred compared to the gNFWOu profile as a best-fit to the Milky Way Galactic rotation curve.

\section{Observational limits}
\label{sec:exp_lim}
We investigate the experimental limits on the photon flux from the Galactic center in a range of photon energies.

At low energies i.e. $E_\gamma \sim 10^{-1} - 50$ GeV, limits on the photon flux are set by the space-based gamma-ray telescope Fermi-LAT. In our analysis, we use the  limits  provided by Ref.~\cite{Picker:2023ybp} where the flux is integrated over an angular region of $\Delta \Omega = 2\pi (1-\cos \theta)$ with $\theta = 1^\circ$ being the angle between the line of sight direction and direction of the Galactic center.

In the energy range $E_\gamma \sim 200 - 5\times 10^4$ GeV, measurements from the ground-based Cherenkov telescope HESS are the most competitive, constraining the observed (differential) photon flux integrated over an annular region with $\theta = 0.5^\circ - 3.0^\circ$ \cite{HESS:2022ygk}. The energy-dependent acceptance of the instrument is obtained from Ref.~\cite{deNaurois:2009ud}.

For heavy $m_\chi \gtrsim 10^{12}$ GeV, most of the flux occurs at higher energies $E_\gamma \gtrsim 10^{4}$ GeV (as can be seen in the bottom left panel of Fig. \ref{fig:1}). In this region, we obtain limits on the angular averaged flux (defined in eq. \eqref{eq:synchrotron_flux_avg}) from the Pierre-Auger Observatory (PAO) \cite{Veberic:2017hwu, Ishiwata:2019aet}. The incrementally improved limits in \cite{Savina:2021cva, PierreAuger:2022uwd, PierreAuger:2022aty} could provide an $O(1)$ improvement \cite{Das:2023wtk} over the constraints presented here.
To speed up the computation of the angular average, we uniformly sample ten points on the unit sphere according to the algorithm laid out in \cite{deserno2004generate} and find the averaged flux over the sampled points. The sample size was optimal as it provided a reliable approximation to samples that included more points, at the price of significantly greater computational cost.

Moreover, CASA-MIA, KASCADE, KASCADE-GRANDE, PAO, TASD and EAS-MSU all provide limits on the (energy) integrated (angular averaged) flux in eq. \eqref{eq:synchrotron_flux_int} \cite{Chianese:2021jke}. Henceforth we will collectively refer to these limits as `Chianese et al.'.

To obtain constraints on the DM lifetime, we require that the photon flux (either synchrotron or prompt) does not exceed the flux measured by the above experiments over the respective energy ranges. We thus very conservatively neglect any photon source besides DM decay (in reality part or most of the emission has other astrophysical origins). 

\begin{figure}[t]
\centering
\includegraphics[width=0.46\linewidth]{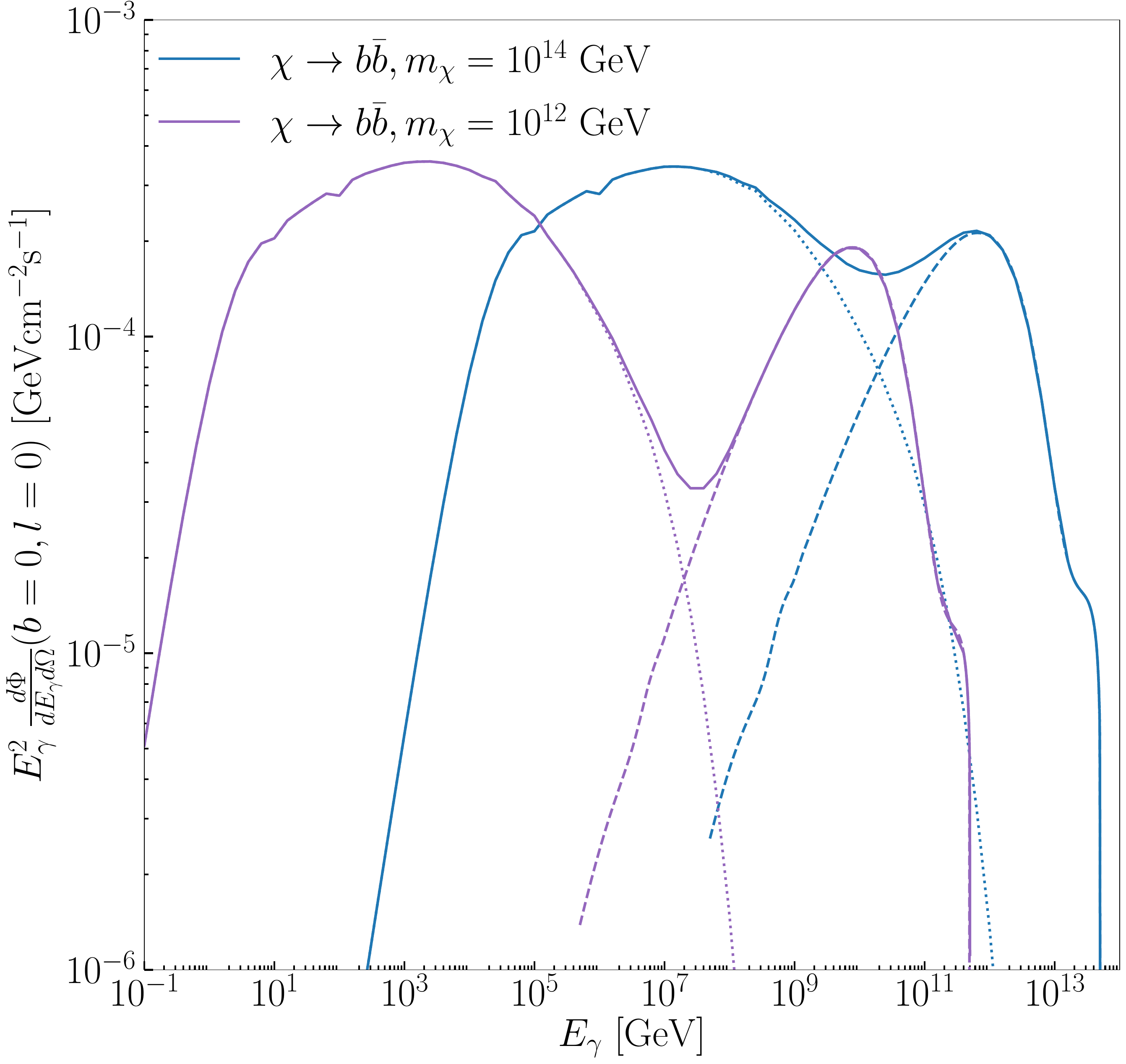} ~~~~
\includegraphics[width=0.46\linewidth]{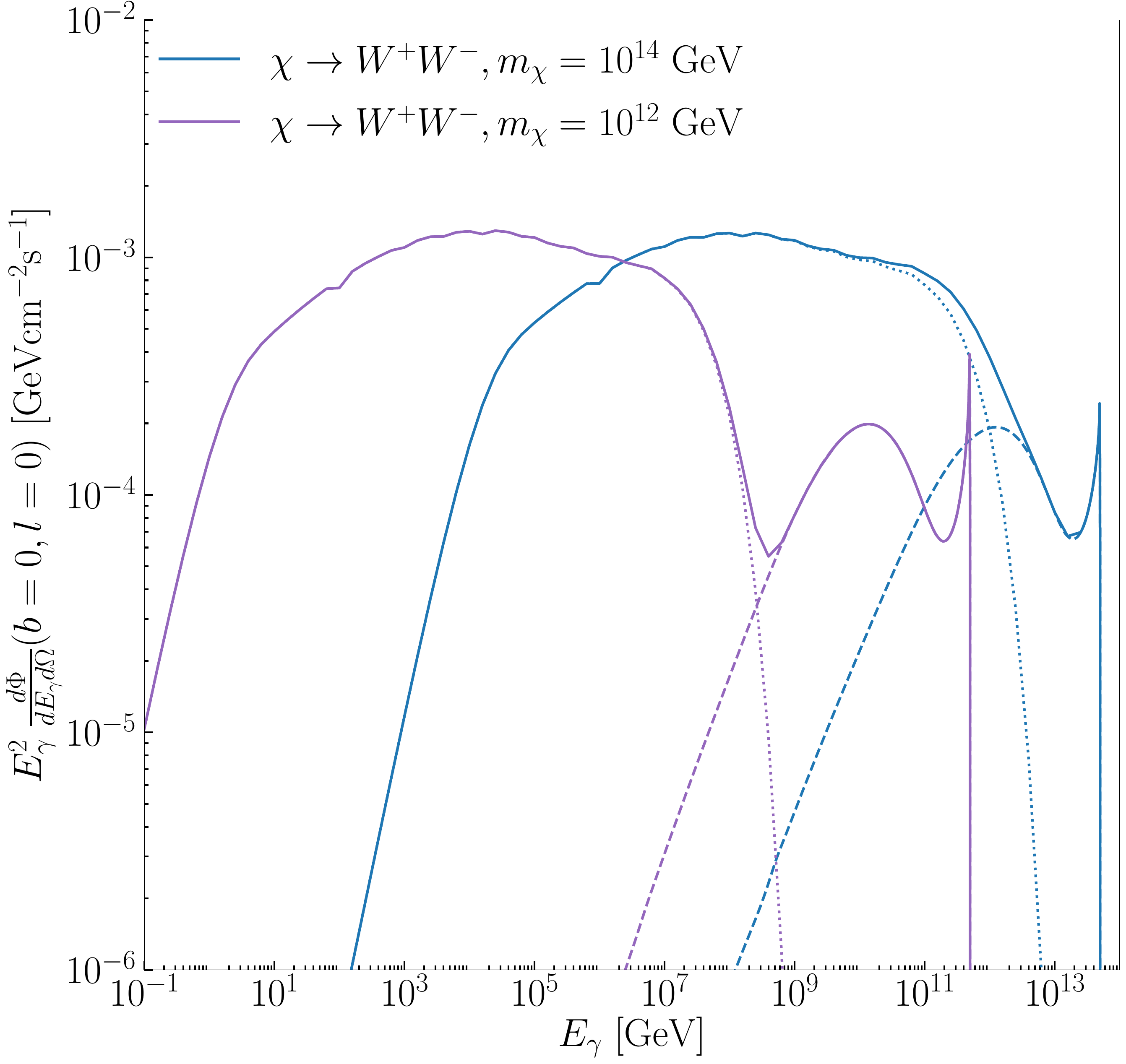}
\caption{Plot of differential photon flux spectra for $b\bar{b}$ (Left) and for $W^+W^-$ (Right) final states. In {\color{c3}blue} is the flux for DM of mass $m_\chi = 10^{14}$ GeV and in {\color{c7}purple} is the flux for DM of mass $m_\chi = 10^{12}$ GeV. The lifetime is fixed to $\tau = 10^{25}$~s. The prompt and synchrotron contributions to the flux are shown as dashed (right bump) and dotted (left bump) lines respectively, while the combined flux is represented by solid lines. } 
\label{fig:spectrum}
\end{figure}

\section{Results} \label{sec:results}
\begin{figure}[tb]
\centering
\includegraphics[width=0.46\linewidth]{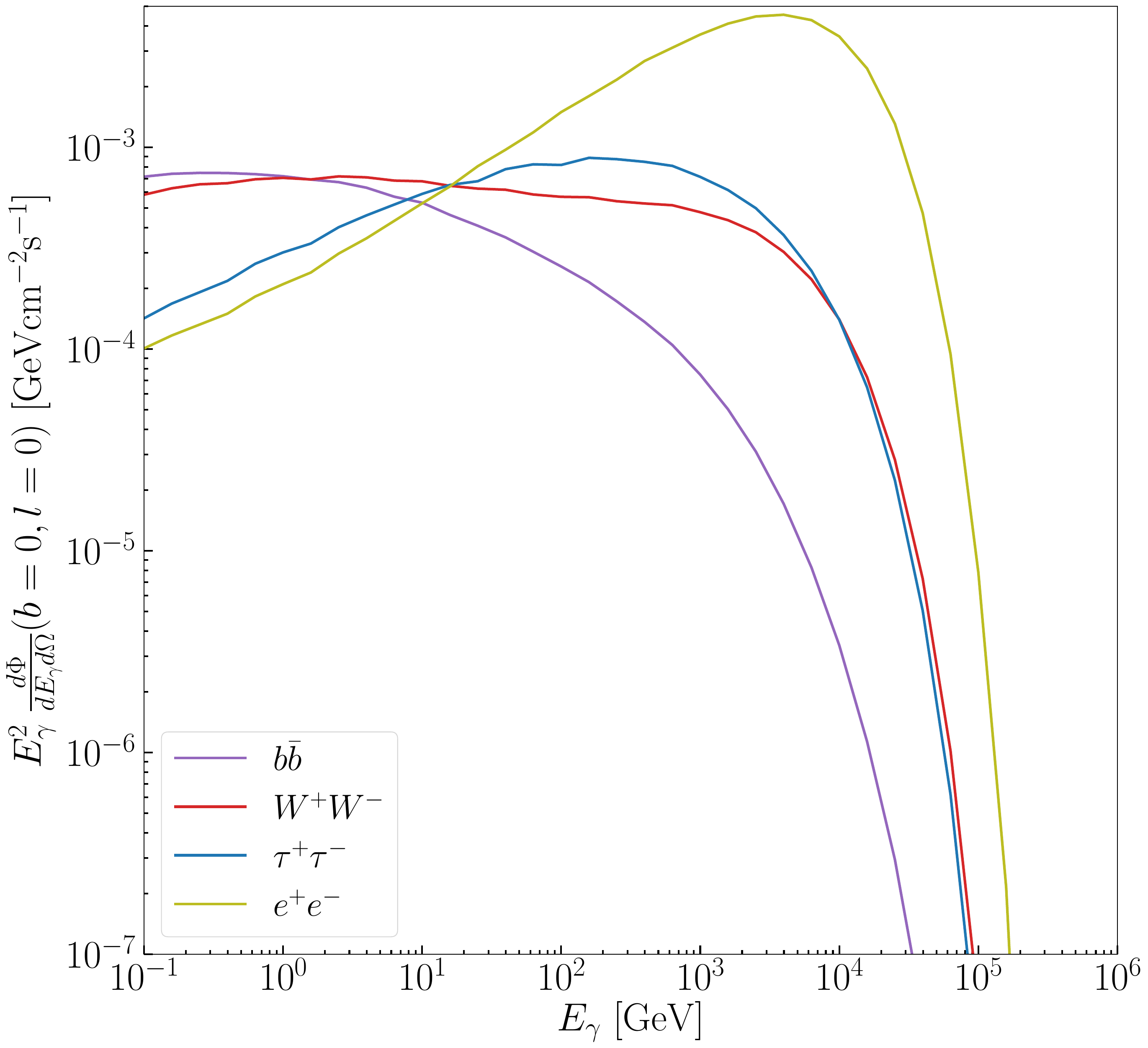} ~~~~
\includegraphics[width=0.46\linewidth]{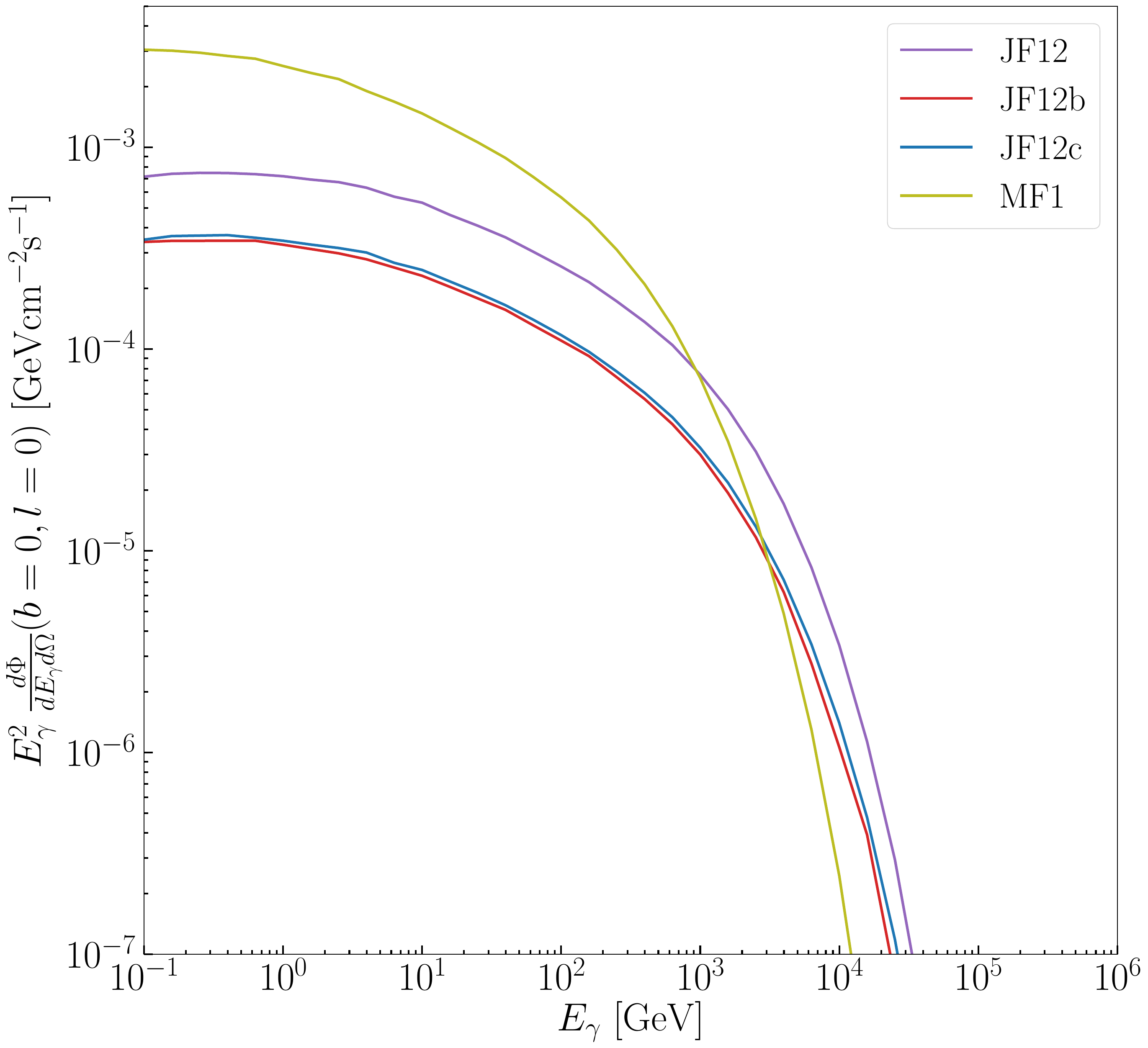}
\includegraphics[width=0.46\linewidth]{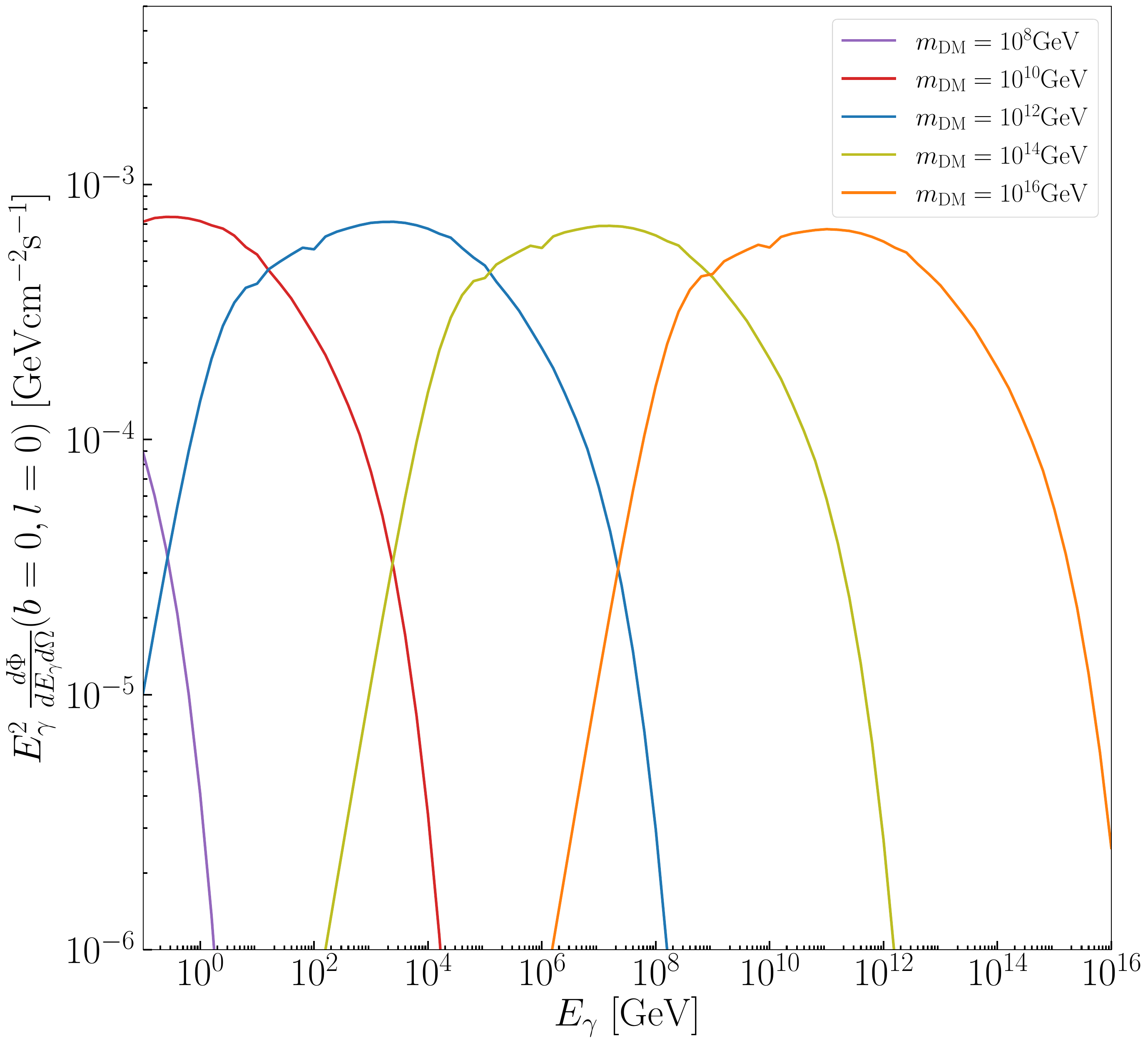}~~~~
\includegraphics[width=0.46\linewidth]{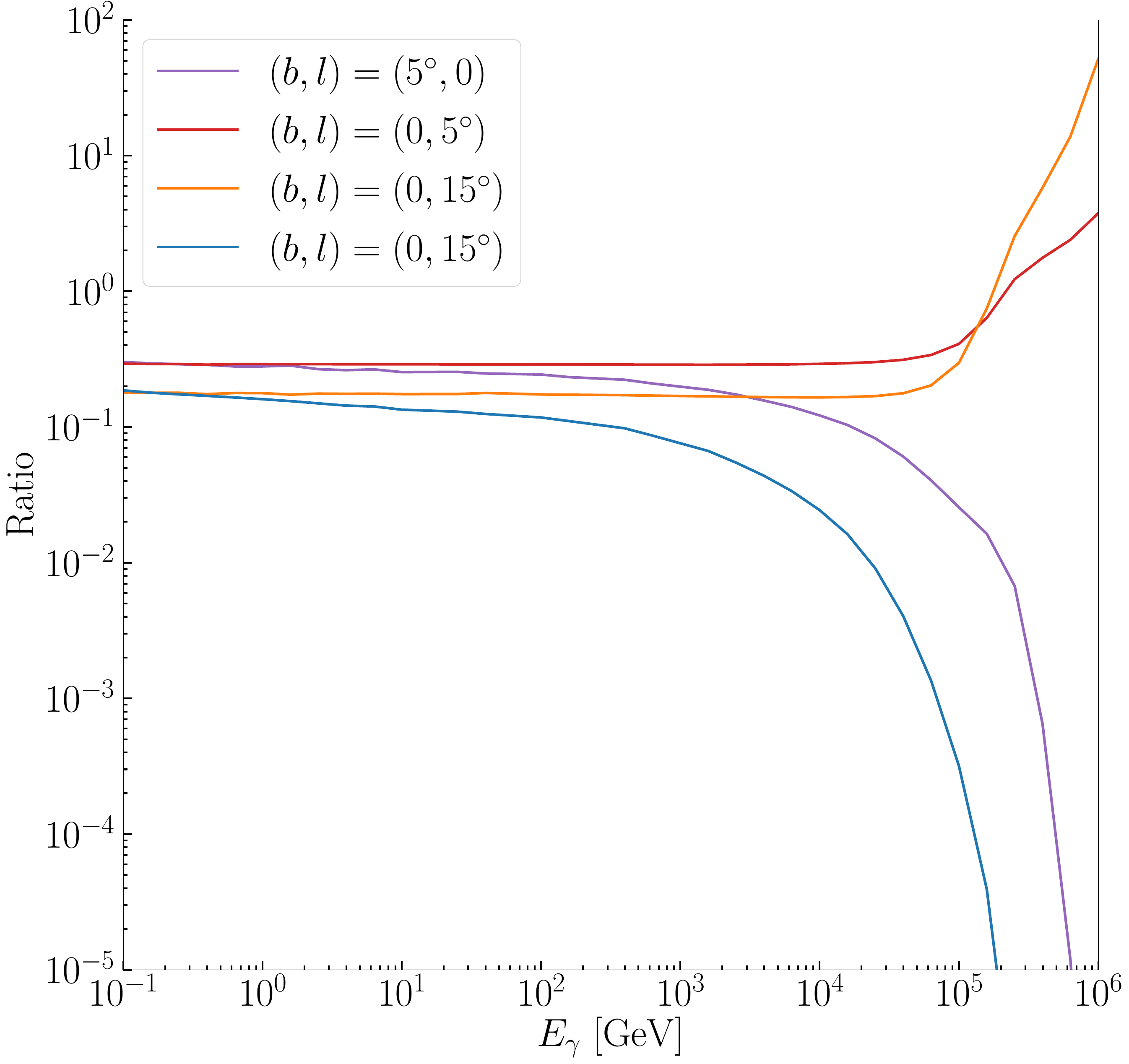}
\caption{The SM initial state, GMF, dark matter mass, and the angular location are taken to be $b\bar{b}$, JF12, $m_\chi = 10^{10}$ GeV and $(b,l) = (0,0)$ by default, with only one of them being varied in each plot whereas the lifetime is fixed to $\tau = 10^{25}$~s. The variation of the differential photon synchrotron flux is depicted for different SM initial states $b\bar{b}$, $W^+ W^-$, $\tau^+ \tau^-$ and $e^+ e^-$ (top left), different magnetic field models JF12, JF12b, JF12c and MF1 (top right), and the range of masses $m_\chi = 10^{8} - 10^{16}$ GeV (bottom left). Finally, the ratio of the differential photon synchrotron flux from different $(b,l)$ locations to the one from $(b,l)=(0,0)$ is shown (Bottom Right).} 
\label{fig:1}
\end{figure}

\begin{figure}[h!]
\centering
\includegraphics[width=0.46\linewidth]{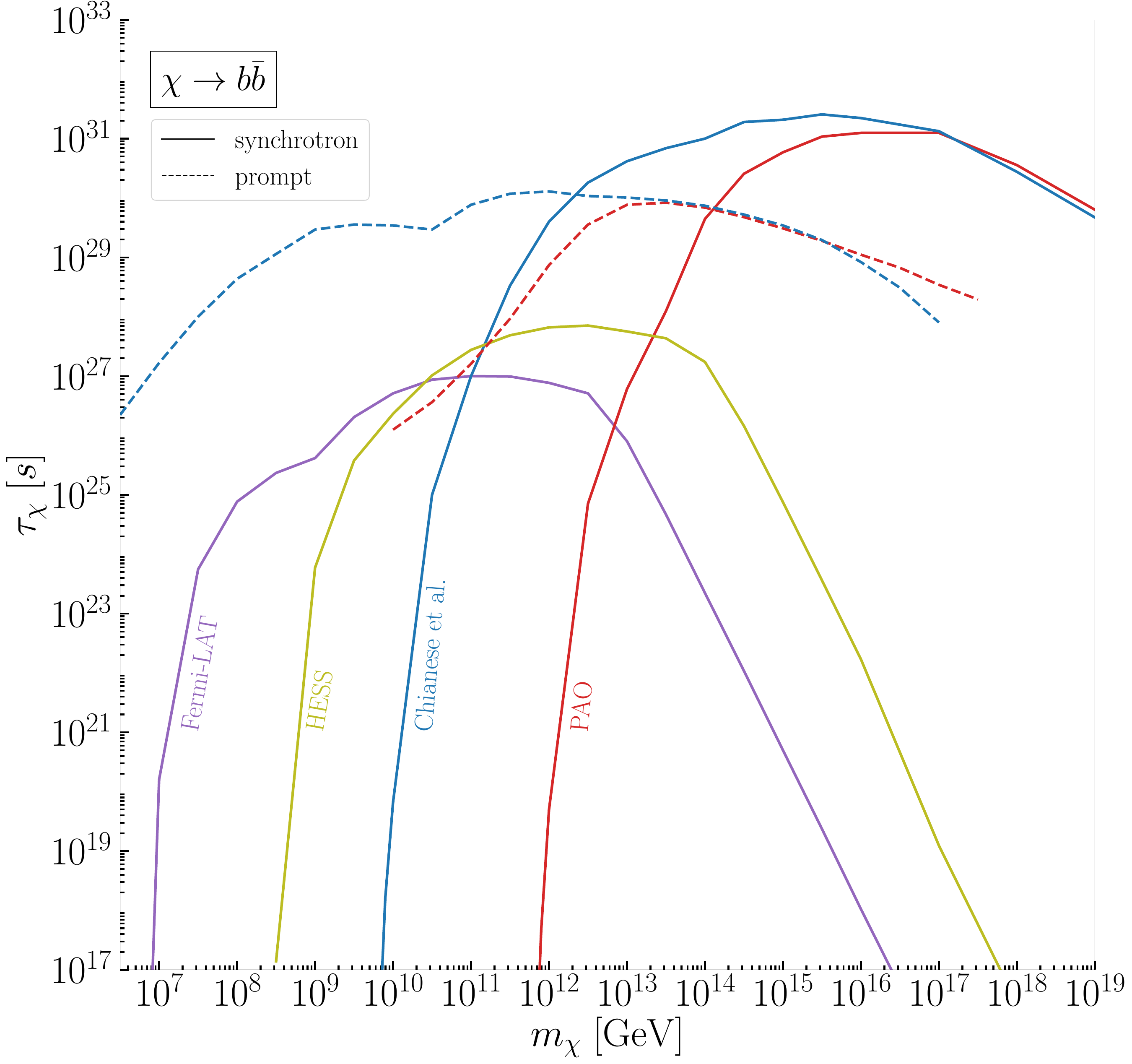} ~~~~
\includegraphics[width=0.46\linewidth]{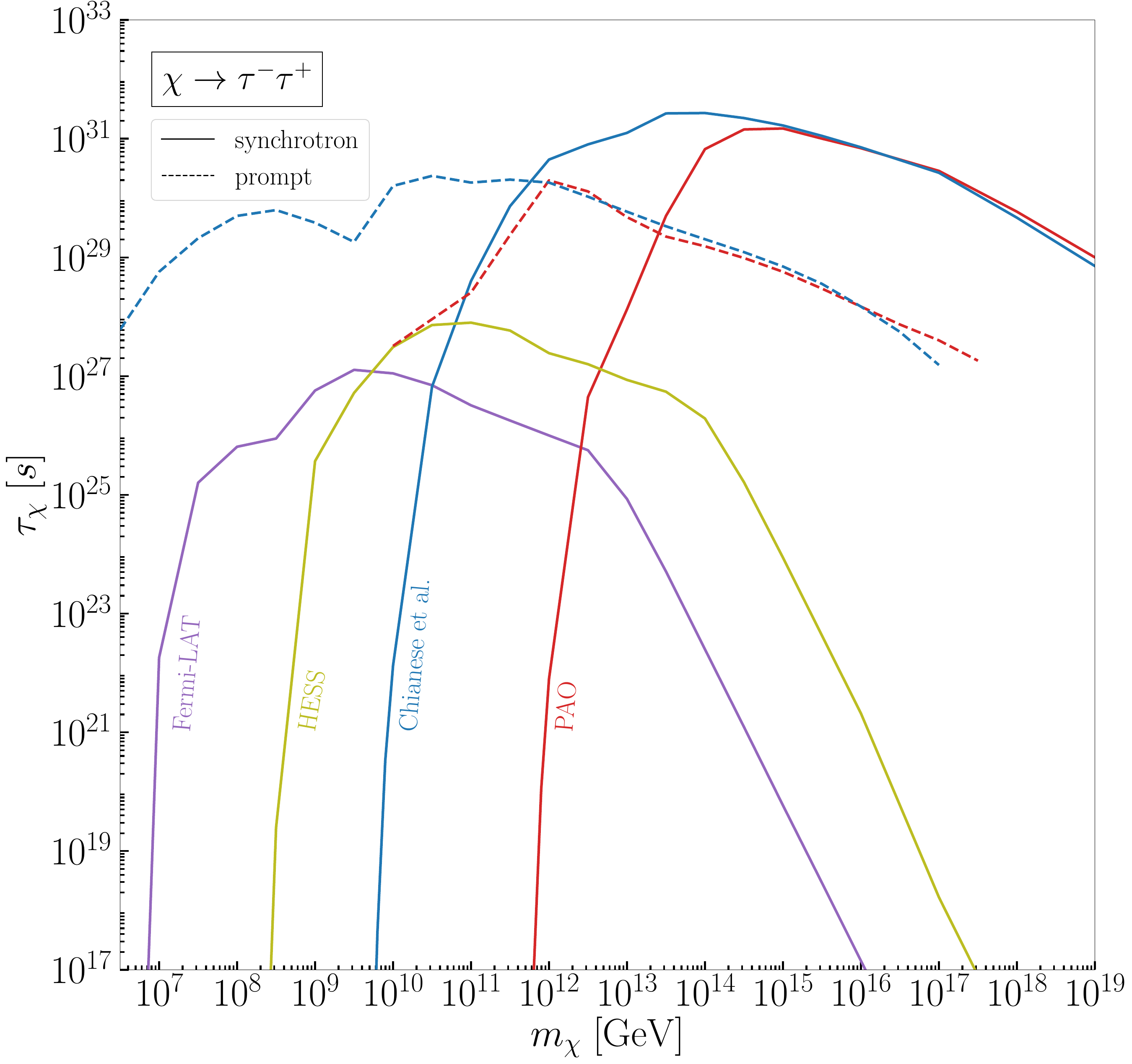}
\includegraphics[width=0.46\linewidth]{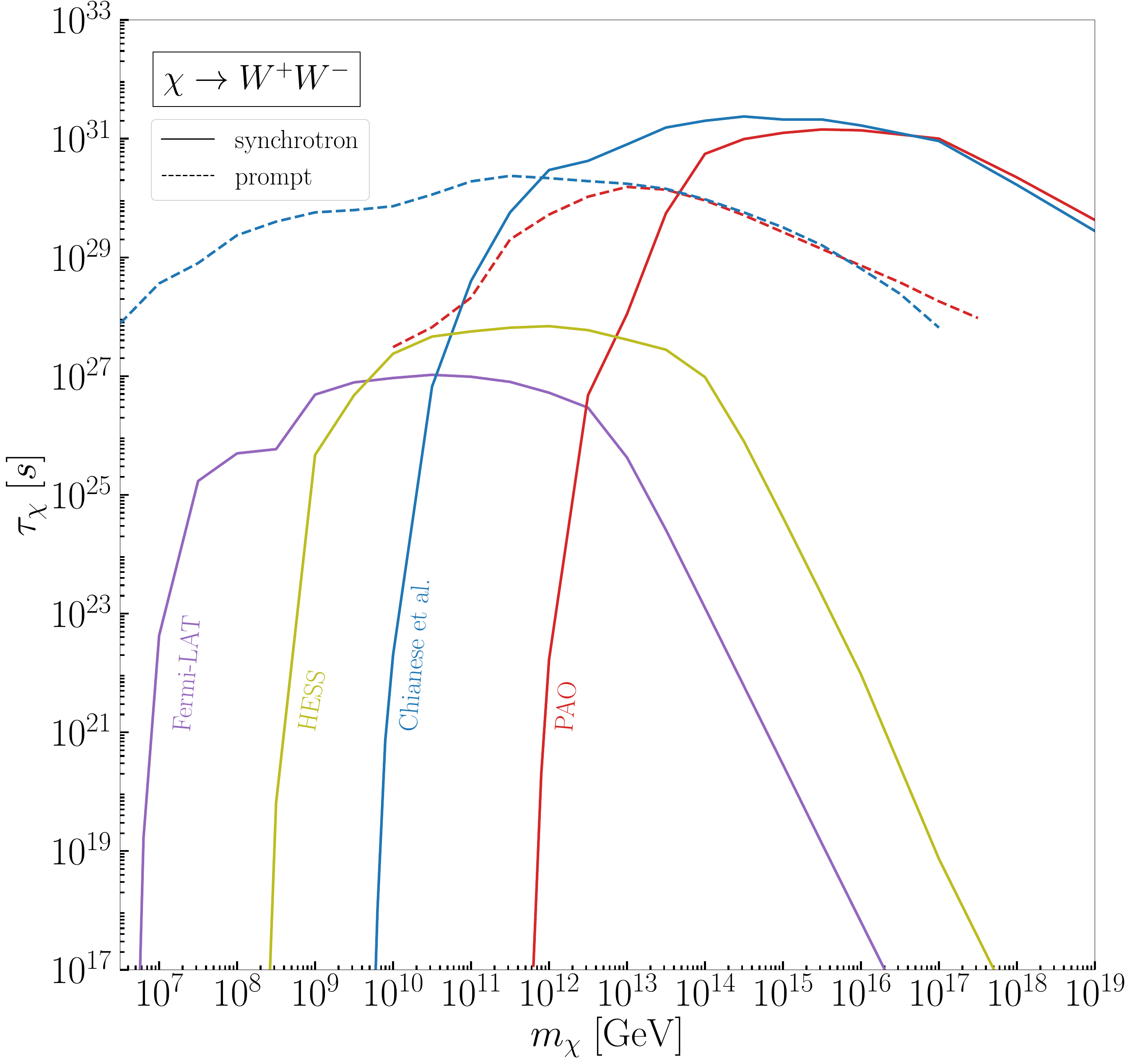}~~~~
\includegraphics[width=0.46\linewidth]{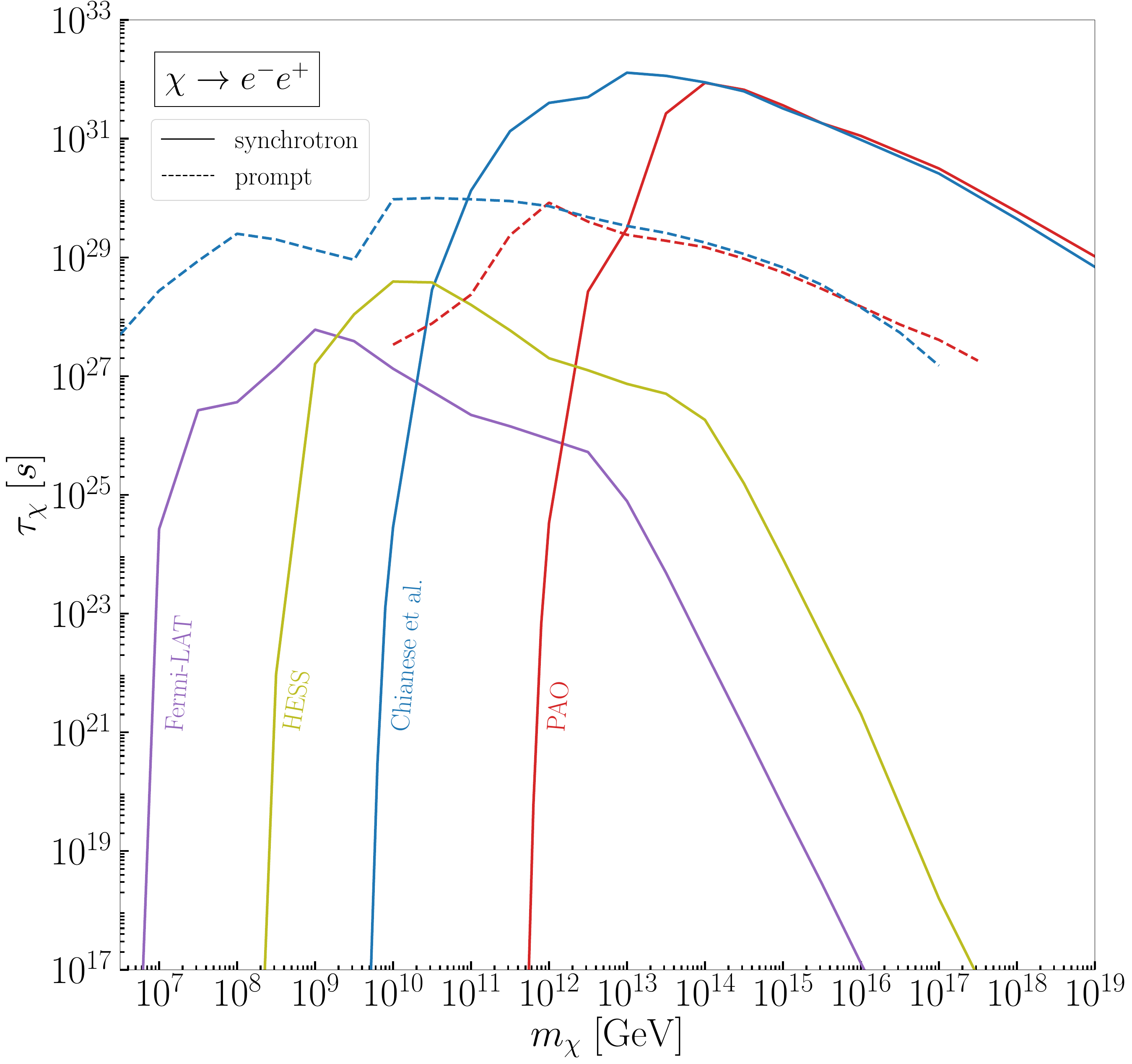}
\caption{The constraints obtained from the different experiments on the lifetime of the DM $\tau_\chi$ as a function of the DM mass $m_\chi$ are shown. The constraints from Fermi-LAT ({\color{c7} purple}), HESS ({\color{c4}green}) and PAO ({\color{c2}red}) are obtained from Refs. \cite{Picker:2023ybp}, \cite{HESS:2022ygk} and \cite{Ishiwata:2019aet} respectively, whereas the constraint from the experiments considered in Ref. \cite{Chianese:2019kyl} are collectively denoted as `Chianese et al.' ({\color{c3} blue}). Each panel shows the constraints obtained for different initial SM states ($b\bar{b}, \tau^+\tau^-, W^+W^-$ and $e^+ e^-$) for both the prompt (dashed lines) and the synchrotron (solid lines) components separately.} 
\label{fig:2}
\end{figure}

\begin{figure}[h!]
\centering
\includegraphics[width=0.46\linewidth]{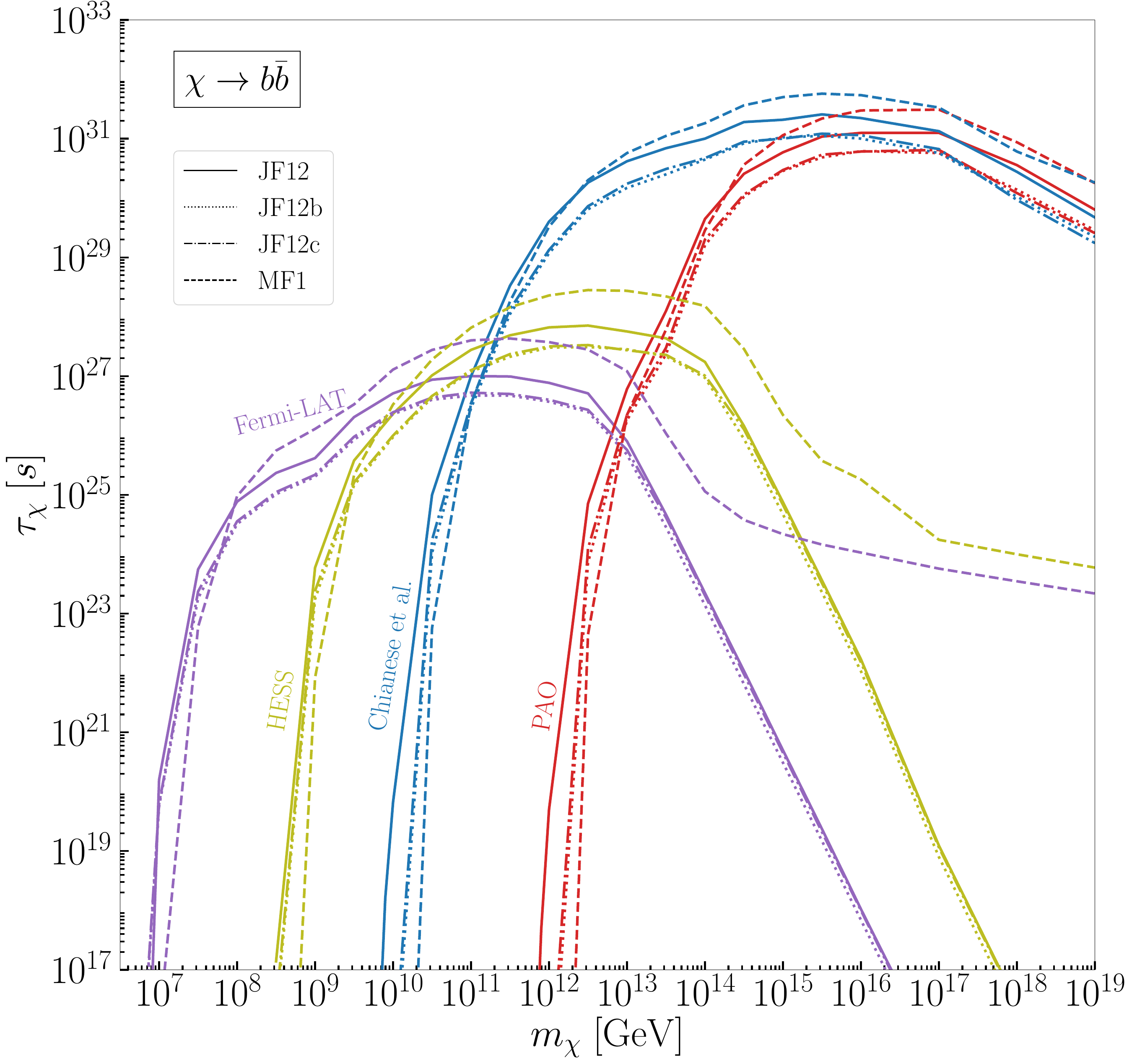} ~~~~
\includegraphics[width=0.46\linewidth]{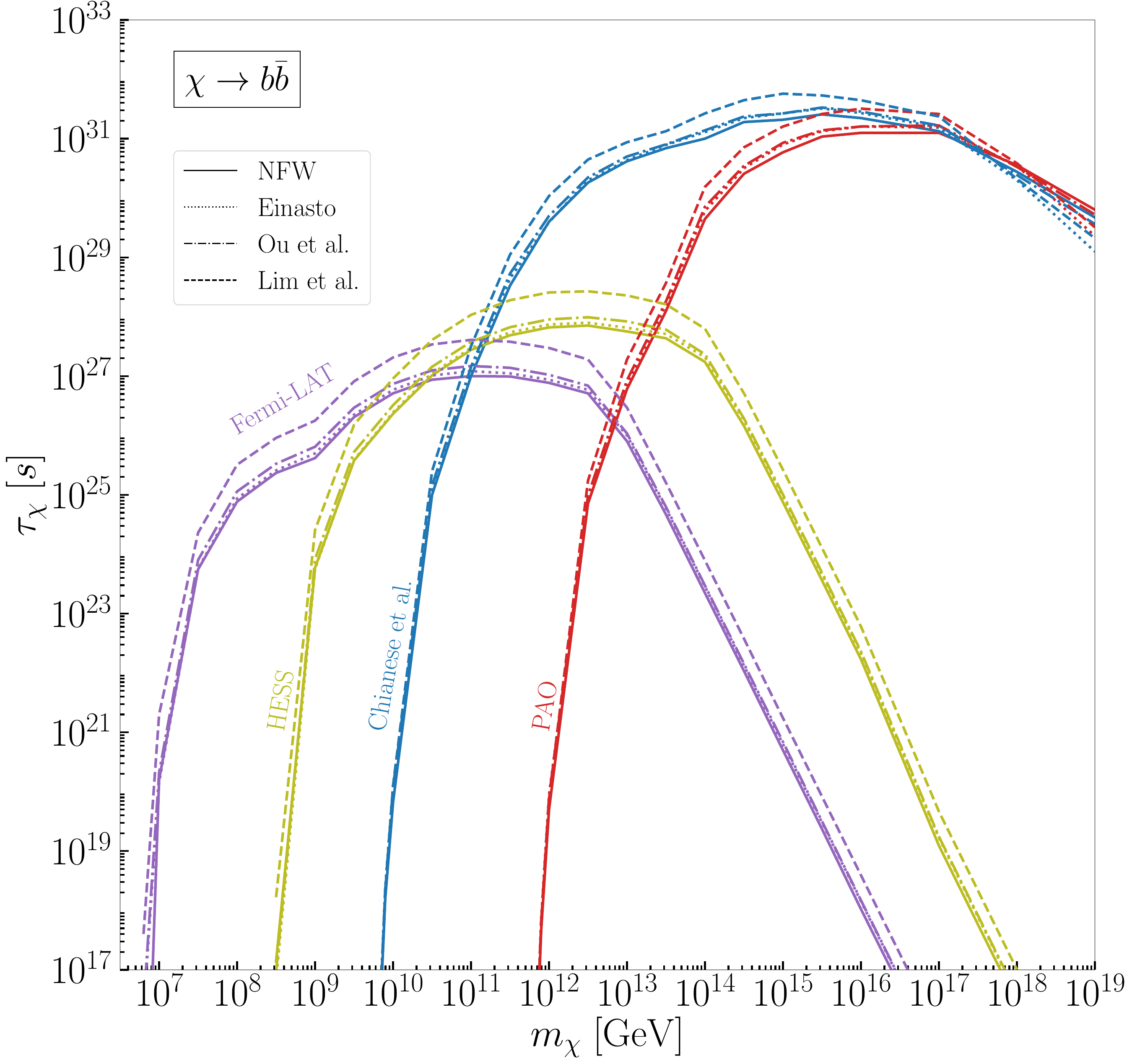}
\caption{The constraints obtained from the different multimessenger experimental probes on the lifetime of the DM $\tau_\chi$ as a function of the DM mass $m_\chi$ are shown. (Left) The Galactic Magnetic Field profile is varied. (Right) The Dark Matter density profile is varied.} 
\label{fig:3}
\end{figure}

We start by showing, in Fig.~\ref{fig:spectrum}, the prompt and synchrotron spectral energy density (SED) emission for decay final states into $b\bar{b}$ (left) and for $W^+W^-$ (right) final states. We show in {\color{c3}blue} results for a DM particle of mass $m_\chi = 10^{14}$ GeV and in {\color{c7}purple}  of mass $m_\chi = 10^{12}$ GeV. The prompt and synchrotron contributions to the flux are shown as dashed (right bump) and dotted (left bump)  respectively, while the combined flux is represented by solid lines. As expected, the flux for the two masses is relatively self similar, but at increasing masses the peak of the synchrotron emission moves to higher energies {\em quadratically} with $m_\chi$. Remarkably, the peak of the synchrotron SED emission surpasses the prompt emission to photons, indicating that the decay releases, eventually, more energy to $e^\pm$ than to prompt photons. Notice that in the $\bar b b$ case, $e^\pm$ stem from charged pion decay whereas photons arise from neutral pion decay, hence the height of the peak is comparable. In the $W^\pm$ case, we note the peak stemming from internal bremsstrahlung at the highest energies; the synchrotron emission is especially bright, including because of the  $W\to e\nu_e$ decay mode.

In Fig. \ref{fig:1}, the synchrotron differential flux $\frac{d\Phi}{dE_\gamma d\Omega}$ is shown as a function of $E_\gamma$. We choose as a benchmark the flux originating from the Galactic center (of Galactic coordinates $(b,l) = (0,0)$), with $m_{\chi} = 10^{10}$ GeV, initial SM state $b\bar{b}$, JF12 magnetic field model and standard NFW DM profile. Each of the four panels shows the change in the flux with respect to one of these inputs being modified.

The top left panel shows the flux for different initial SM states namely $b\bar{b}$, $W^+ W^-$, $\tau^+\tau^-$ and $e^+ e^-$. We find that massive SM states such as the bottom quark and $W$ tend to have a flatter spectrum than lighter states such as the $\tau$ and the electron which have a more peaked spectrum. This can be ascribed to the production mechanisms associated with each final state: bottom quarks produce photons primarily by the decay of byproducts of the hadronization process, eventually leading to neutral pions decaying preferentially to two photons; $e^\pm$ produce copious internal bremsstrahlung photons, whereas $W^\pm$ and $\tau^\pm$ both produce photons from internal bremsstrahlung and from hadronic channels. 

The top right panel shows the variation of the flux with the different magnetic field configurations: MF1, JF12, JF12b and JF12c, all discussed and detailed upon in Section \ref{subsec:mag_field} above. Notice that JF12b and JF12c are extremely close, and JF12 enhances the synchrotron emission by a factor of around 2, indicating that the magnetic field is within 50\% of the JF12b and JF12c models; the only qualitative outlier is the MF1 model, which misses a number of features such as halo field and turbulent field components included in the other models. Most conspicuously the synchrotron emission computed with the MF1 model has a much softer spectrum that over-shoots that from the JF models by almost one order of magnitude at low energy and under-shoots by similar amounts at high energy. It is relevant to mention that this behavior scales strongly with the DM mass: for example, for $m_\chi = 10^{18}$ GeV, the synchrotron flux from MF1 dominates at low photon energies by up to ten orders of magnitude over that from JF12, implying much stronger constraints from Fermi-LAT and HESS in this region, as can be seen in Fig. \ref{fig:3}. Nevertheless, we reiterate that due to the crudeness of MF1, the constraints there are not expected to be physical, as they deviate too far from the more reliable and complete JF12 models.

The bottom left panel shows the flux for a range of DM masses from $m_{\chi} = 10^8$ to  $10^{16}$ GeV. As the DM mass increases, the peak of the spectrum also appears at higher photon energies, as expected. As indicated in the Introduction, we find that the peak emission occurs at $E_{\rm peak}/{\rm GeV}\simeq (m_\chi/(10^{10}\ {\rm GeV}))^2$. 

Finally, the bottom right panel shows the angular variation of the flux at different Galactic angular coordinates $(b,l) = (5^\circ,0)$, $(b,l) = (0,5^\circ)$, $(b,l) = (15^\circ,0)$ and $(b,l) = (0,15^\circ)$ as compared to the flux from the Galactic center (GC). As expected, the GC direction dominates at all energy, mostly because of a higher DM density along the line of sight. However, interestingly enough, at higher Galactic longitudes, along the Galactic plane ($b=0$), we find a larger emission at high energy than in the GC direction: this is presumably due to a complex combination of effects related to the injected electron equilibrium spectrum and the magnetic field along the line of sight compensating and out-doing the smaller DM density.

In Fig. \ref{fig:2}, we derive constraints on the DM lifetime $\tau_\chi$ as a function of its mass $m_\chi$, and compare with the corresponding constraints from prompt emission. Each panel shows the bounds obtained for the different initial SM states $b\bar{b}, W^+W^-, \tau^+\tau^-, e^+ e^-$. The {\color{c2}red}, {\color{c3} blue}, {\color{c7} purple}, {\color{c4} green} lines show the constraints derived from the Pierre-Auger observatory \cite{Ishiwata:2019aet}, `Chianese et al.' \cite{Chianese:2021jke}, Fermi-LAT~\cite{Picker:2023ybp} and HESS~\cite{HESS:2022ygk} respectively, with the solid lines corresponding to the synchrotron emission and the dashed lines from the prompt emission (which, in the range of masses under consideration corresponds to the Chianese et al and PAO limits). We note that for all final states, the synchrotron component provides a significant improvement over the constraints coming from the prompt component only for $m_{\chi} \gtrsim 10^{12}\ {\rm GeV}$ by up to, and in some cases over one order of magnitude, while the prompt emission is more constraining, by a similar amount, up to one order of magnitude at lower masses $m_{\chi} \lesssim 10^{12}\ {\rm GeV}$.

In Fig. \ref{fig:3}, we derive the same type of constraints from the synchrotron emission, but varying the {\em astrophysical inputs} for the magnetic field models (Left) and for different DM profiles (Right). In the left panel, the solid, dashed, dot-dashed and dotted lines represent the bounds from JF12, MF1, JF12c and JF12b respectively whereas in the right panel, they represent the bounds from NFW, gNFWLim, gNFWOu and Ein respectively. With the caveat discussed above about MF1, we find that our predictions are broadly subject to up to one order of magnitude uncertainty from the magnetic field model, and slightly less than that for the dark matter density. Note that the prompt emission is also subject to the latter range of uncertainty, but it is  unaffected by the magnetic field. However, one should also note that HDMSpectra has considerable uncertainties when $x = 2 E_e/m_\chi$ is small, notably $x \lesssim 10^{-4}$. Hence we restrict $x\geq 10^{-4}$ for the synchrotron component and $x\geq10^{-6}$ for the prompt component\footnote{Since the prompt flux vanishes entirely below the cut-off contrary to the synchrotron component, we chose the minimum $x$ value allowed by {\tt HDMSpectra} for the prompt component.}.

\section{Conclusions and Outlook} \label{sec:conclusions}
We explored how gamma-ray telescopes such as Fermi-LAT and HESS, and high-energy observatories such as Pierre Auger, provide the most stringent constraints on the lifetime of heavy DM candidates, specifically heavier than $10^{12}$ GeV, by constraining the synchrotron emission generated by the very high-energy electrons and positrons produced in the decay, rather than the prompt gamma-ray emission. The synchrotron constraints are up to one order of magnitude stronger than constraints from the prompt emission by very high-energy cosmic-ray and gamma-ray facilities.

The synchrotron luminosity depends on the specific Galactic magnetic field model assumed; we showed, however, that for realistic, detailed magnetic field models the uncertainty is well below one order of magnitude, and comparable to the uncertainty associated with the dark matter density profile.

We showed that the synchrotron emission peaks at an energy  $E_{\rm peak}/{\rm GeV}\simeq (m_\chi/(10^{10}\ {\rm GeV}))^2$, and is typically brighter in the direction of the Galactic center, i.e. for $(b,l)=(0,0)$, with the exception of very high energies, where it can be brighter at non-zero longitude along the Galactic plane, possibly offering a better signal-to-noise ratio as there the astrophysical backgrounds are typically lower.

In the future we plan to re-consider, with the present implementation of several up-to-date Galactic magnetic field models, limits from synchrotron emission of lighter DM candidates. We also plan to assess how information on polarization \cite{Manconi:2022vci} could improve on the limits presented here, should future gamma-ray telescopes be sensitive to a polarized signal. Finally, the tools developed in this study will be made available by request to the Authors, and can be used to set constraints on concrete model-specific DM candidates.

\section*{Acknowledgements} 
We are very grateful to Noémie Globus for helpful discussions concerning the JF12 Galactic magnetic field. We also thank Tracy Slatyer for bringing triplet pair production processes to our attention, and Kohta Murase, Tarak Nath Maity for helpful comments. This work is partly supported by the U.S. Department of Energy, grant number de-sc0010107.

\begin{appendix}
\section{synchrotron power} \label{app:Psyn}
The synchrotron power is calculated using \cite{Colafrancesco:2005ji,Buch:2015iya,Profumo:2010ya}
\begin{equation}
\label{eq:Psyn}
    P_{\rm syn} (E_\gamma, E_e, s, b, l) = \frac{\sqrt{3}}{2}\frac{e^3}{m_e}\int_0^\pi d\theta ~ B \sin^2\theta \, F \left ( \frac{x}{\sin\theta}\right ),
\end{equation}
where $e$ is the electric charge and $B$ is the magnetic field. The other functions featuring in the integrand are 
\begin{align}
    x &= \frac{E_\gamma}{3\pi \nu_0 \gamma^2 } \left [ 1 + \left ( \gamma \frac{2\pi\nu_p}{E_\gamma}\right )^2 \right ]^{3/2}\\
    F(t) &= t \int_t^\infty dz\, K_{5/3}(z)
\end{align}
where $\gamma = \frac{E_e}{m_e}$ is the Lorentz boost factor, $\nu_p \simeq 8980 \text{Hz} \left (\frac{n}{ \text{cm}^{-3}}\right )^{1/2}$ is the plasma frequency, $\nu_0 = (eB)/(2\pi m_e)$ is the gyro-frequency, and $n \simeq 1 \text{cm}^{-3}$ is the thermal electron number density.


\section{Energy losses}
\label{app:energy_losses}
$e^\pm$ at high energies lose energy mainly through Inverse Compton (IC) scattering and triplet pair production processes (TPP) against CMB, SL and IR photons, as well as synchrotron radiation from being accelerated by the Galactic magnetic field.
\begin{equation}
    b_{\rm tot} \simeq b_{\rm syn} + b_{\rm IC} + b_{\rm tpp}
\end{equation}

The synchrotron energy loss is given by
\begin{equation}
    b_{\rm syn}(E_e) = \frac{4\sigma_T E_e^2}{3 m_e^2}\frac{B^2}{2}
\end{equation}
where $\sigma_T$ is the Thomson cross-section
\begin{equation}
\sigma_T = \frac{8\pi\alpha_{\rm em}^2}{3m_e^2}
\end{equation}
with $\alpha_{\rm em}$ being the fine-structure constant.

The number density of photons (taking only the CMB component) at an energy $\epsilon$ is 
\begin{equation}
    n(\epsilon) = \frac{\epsilon^2}{\pi^2}\frac{1}{e^{\epsilon/T_{\rm CMB}}-1}
\end{equation}
where we neglect the starlight and dust contributions. This is an excellent approximation since for highly energetic electrons (as typically produced from superheavy DM) with an energy above $E_e \gtrsim 50 $ TeV, scattering with higher energy photons from SL and IR backgrounds are more Klein-Nishina suppressed \cite{Sudoh:2021avj, Cirelli:2010xx}, and therefore contribute negligibly to the IC energy loss compared to the CMB.

The IC energy loss is obtained from the following expression
\begin{equation}
    b_{\rm IC}(E_e) = 3 \sigma_T \int_0^{\infty} d\epsilon \int_{1/(4\gamma^2)}^1 dq\, \epsilon \, n(\epsilon) \frac{(4 \gamma^2 - \Gamma)q-1 }{(1+\Gamma q)^3} \left [ 2q \log q + q + 1 - 2 q^2 + \frac{\Gamma^2 q^2 (1-q)}{2 (1+\Gamma q)} \right ]
\end{equation}
where $\gamma = E_e/m_e$ is the Lorentz boost of the electron and $\Gamma(E_e,\epsilon) = 4\epsilon\gamma/m_e$.

Triplet pair production (TPP) $\gamma e \to 3 e$ is an $O(\alpha_{\rm em}^3)$ QED process where a photon interacts with an electron/positron to produce an electron-positron pair in addition to the original recoiling electron/positron. For highly energetic electrons, the energy loss by TPP becomes comparable to that of IC.

In the extreme Klein-Nishina limit ($\Gamma \gg 1$), Ref.~\cite{10.1093/mnras/253.2.235} derives an analytical estimate of the energy loss valid in the regime $\gamma\epsilon \gtrsim 10^3 \, m_e$,
\begin{equation}
b_{\rm tpp} (E_e) \simeq \frac{15}{8}\sigma_T m_e^2 \int d\epsilon \frac{n(\epsilon)}{\epsilon} \alpha_{\rm em} \left ( \frac{\gamma \epsilon}{m_e} \right )^{1/4} \left [ \frac{28}{9}\ln \left (\frac{2\gamma \epsilon}{m_e} \right ) - \frac{218}{27} \right ]~.
\end{equation}

\end{appendix}


\bibliography{bibliography}

\end{document}